\documentclass[review]{elsarticle}

\usepackage[numbers]{natbib}

\usepackage{url} 
\usepackage{graphicx}

\title{A physical perspective on lithium therapy}
\author[1]{Dana Kamp}
\ead{dana.taylor.kamp@nbi.ku.dk}
\affiliation[1]{organization={The Niels Bohr Institute},
	city={Copenhagen},
	country={Denmark}}

\begin{document}

\begin{abstract}
Lithium salts have strong medical properties in neurological disorders such as bipolar disorder and lithium-responsive headaches \cite{pereira_da_silva_neto_lithium-responsive_2010, lindner_hypnic_2023, silva-neto_hypnic_2019}. They have recently gathered attention due to their potential preventive effect in viral infections \cite{ferensztajn-rochowiak_long-term_2023, rybakowski_antiviral_2022}. Though the therapeutic effect of lithium was documented by Cade in the late 1940s, its underlying mechanism of action is still disputed \cite{malhi_therapeutic_2016}. Acute lithium exposure has an activating effect on excitable organic tissue  and organisms, and is highly toxic \cite{schou_biology_1957, butler-munro_lithium_2010, izsak_differential_2021, yip_lithium_2007}.  
Lithium exposure is associated with a strong metabolic response in the organism, with large changes in phospholipid and cholesterol expression \cite{lopez-corcuera_change_1988,pettegrew_effects_2001, aliyazicioglu_treatment_2007}. Opposite to acute exposure, this metabolic response alleviates excessive cellular activity \cite{mertens_differential_2015, stern_neurons_2018, santos_deficient_2021}.\\
\indent The presence of lithium ions strongly affects lipid conformation and membrane phase unlike other alkali ions \cite{hauser_interactions_1983}, with consequences for membrane permeability, buffer property and excitability. This review investigates how lithium ions affect lipid membrane composition and function, and how lithium response might in fact be the body's attempt to counteract the physical presence of lithium ions at cell level. Ideas for further research in microbiology and drug development are discussed.

\end{abstract}

\maketitle

\section{What is lithium therapy?}

	\subsection{Medical use}
	Lithium therapy is a prophylactic treatment administered in the form of a salt, usually taken orally on a daily basis. The uptake through the gastrointestinal tract is highly efficient, with a 80 - 100\% bio-availability \cite{grandjean_lithium_2009} and blood serum levels reach a maximum after one to two hours \cite{grandjean_lithium_2009}. Lithium is distributed quite evenly in the aqueous phase of the body, compared to the more biologically prevalent alkali ions, sodium and potassium, that have actively upheld concentration gradients. Still, there are measurable differences in concentration and rates of accumulation of lithium ions at the tissue level and cellular level \cite{schou_biology_1957}. \\

Lithium is detectable in soft tissue, bone and in the extracellular space of the brain within minutes of ingestion \cite{schou_biology_1957, hillert_uptake_2012}, though the cellular uptake and clearance is delayed in the central nervous system compared to other tissue types \cite{schou_biology_1957, hillert_uptake_2012}. Generally, lithium concentration in the brain is estimated to vary within 50-80\% of serum levels \cite{grandjean_lithium_2009} and to accumulate to a stable level over the course of weeks. In rat models the intracellular lithium concentration matches serum concentrations 6 hours after acute administration \cite{hillert_uptake_2012}. \\

Lithium can partially substitute sodium in biological functions such as neuronal- and muscular excitability. It is possible to swap up to 40-50\% of extracellular sodium with lithium in tissue samples \textit{in vitro} and preserve excitability \cite{keynes_effect_1959, keynes_permeability_1959}, though the excitability will be highly modulated. Excitability is sometimes retained at a 100 \% substitution, but high levels of lithium are associated with long term toxicity effects \cite{schou_biology_1957}. \\

	\noindent\textit{General use and concentrations} \\
Lithium is a famous first-choice treatment for bipolar disorder \cite{malhi_therapeutic_2016, gao_chapter_2021, ferensztajn-rochowiak_ultra-long-term_2021, kessing_observational_2012}, a spectrum disorder characterized by episodal disturbances in mood and activity levels with a general population prevalence of 3-5\% \cite{angst_emerging_1998, rybakowski_response_2014}. It has also found use in subpopulations of patients diagnosed with unipolar depression \cite{johnson_historical_1975, bschor_lithium_2014}, schizophrenia \cite{de_sousa_early_2012, luo_lithium_2020} and neurodegenerative diseases \cite{britti_mitochondrial_2018, matsunaga_lithium_2015}. Less known is its effectiveness in pain management of a lithium-responsive subclass of primary headaches \cite{pereira_da_silva_neto_lithium-responsive_2010, lindner_hypnic_2023, silva-neto_hypnic_2019}. A few studies have indicated chronic lithium treatment suppresses outbreaks of herpes simplex virus \cite{ferensztajn-rochowiak_long-term_2023, rybakowski_antiviral_2022}, that coincidentally reside in nerve tissue \cite{murru_lithiums_2020}.  Similar for all illnesses is an optimal response at a stable blood concentration slightly less than toxicity, which occurs at serum levels around 1.5 mM  \cite{grandjean_lithium_2009}. Therapeutic effect is obtained at blood concentrations of 0.6-0.8 mM for mental symptoms (Often higher doses in acute mania) \cite{grandjean_lithium_2009, hsu_differences_2022} and 0.4-0.8 M for cluster headaches \cite{pereira_da_silva_neto_lithium-responsive_2010, lindner_hypnic_2023}. Supression of herpes outbreaks have been observed at subclinical concentrations of $<$ 0.56 mM \cite{amsterdam_suppression_1996}.\\

\noindent\textit{Lithium responsiveness}\\
It is a general observation that the response to lithium treatment is not uniform within the bipolar patient population. This tendency has caused researchers to divide the patient population into various degrees of lithium response, with the extremes 'excellent lithium responders' and 'lithium non-responders' taking up each end of the spectrum, and various definitions of semi-responding groups in between \cite{alda_who_2017}. Excellent lithium response is defined as a complete remission of symptoms under long-term lithium monotherapy, while lithium non-response is an absence of therapeutic advantage, often with considerable negative effects \cite{reisberg_side_1979}. The gradual response to lithium is the primary entry point when investigating the underlying mechanism of lithium treatment and the pathology of mental illness.\\

\noindent\textit{Excellent lithium response and daily rhythm}\\
Excellent lithium response is well-documented in bipolar disorder, and estimated to make up between 6\% - 30\% of the patient population \cite{kessing_predictors_2011, rybakowski_prophylactic_2001}. In these patients, long-term lithium treatment will not only attenuate future mood fluctuations, but also reduce some of the risks associated with the illness, such as cancer, dementia, and cardio- and cerebrovascular diseases \cite{jakobsson_towards_2017, kessing_increased_2003, ponzer_lithium_2023, chen_cardioprotective_2023}. This ability is unparalleled compared to other pharmacological therapies \cite{rybakowski_antiviral_2022}. It indicates that lithium not only ameliorates symptoms, it also treats an underlying pathology shared by this particular subgroup of the bipolar population \cite{alda_who_2017}. \\
\indent Excellent lithium response is associated with a distinct set of symptoms \cite{abou-saleh_who_1993, amare_association_2021}. Patients are often 'early birds' \cite{rohr_impact_2022}, and their symptomatic patterns are clear episodes of mania followed by depression \cite{schou_biology_1957, kleindienst_which_2005, alda_who_2017}. In this subgroup, there is a high heritability of both illness and lithium response \cite{grof_is_2002, alda_who_2017}. Similarly, lithium-responsive headaches are known for their rhythmicity, specifically 'morningness', seasonal pattern, and heritability \cite{costa_preliminary_2015, lindner_hypnic_2023, pereira_da_silva_neto_lithium-responsive_2010}, but little data exists on the diversity of lithium response within this patient group. Interestingly, this improvement in diurnal rhythm has, for bipolar lithium responders, been linked to metabolic changes down to the cellular level.\\

	\subsection{Investigating lithium response}
\begin{figure}
	\centering
	\includegraphics[width=0.7\linewidth]{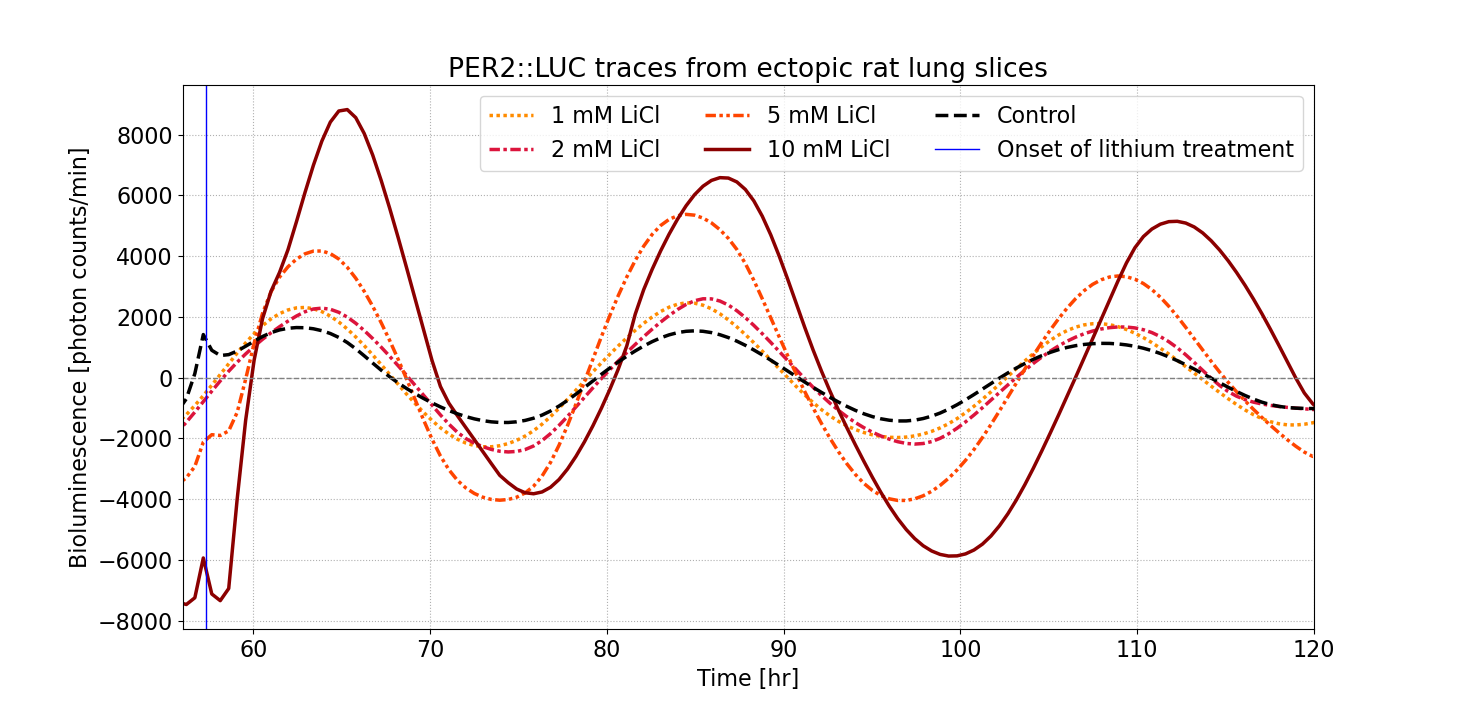}

	\vspace{10mm}
	\begin{minipage}{0.45\textwidth}
		\includegraphics[width=0.8\linewidth]{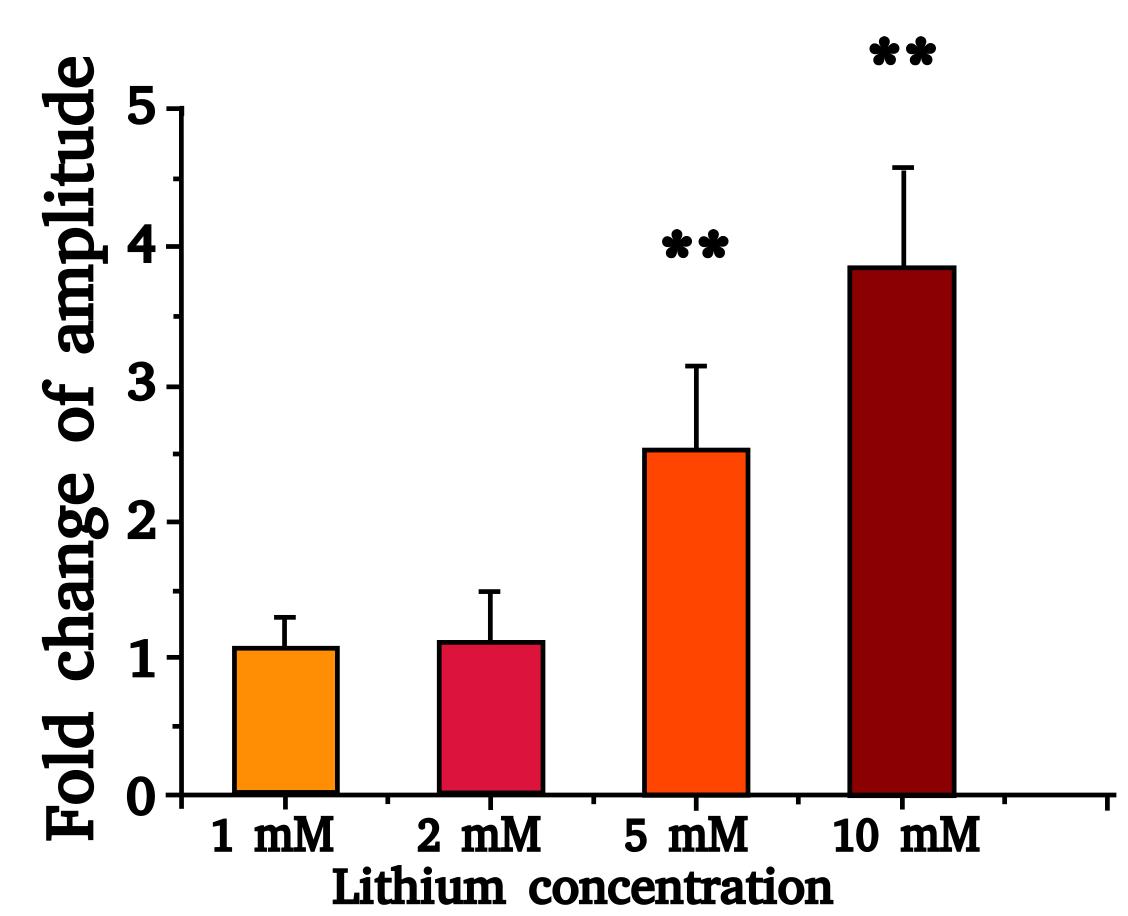}
	\end{minipage}
	\hspace{5mm}
	\begin{minipage}{0.45\textwidth}
		\includegraphics[width=0.8\linewidth]{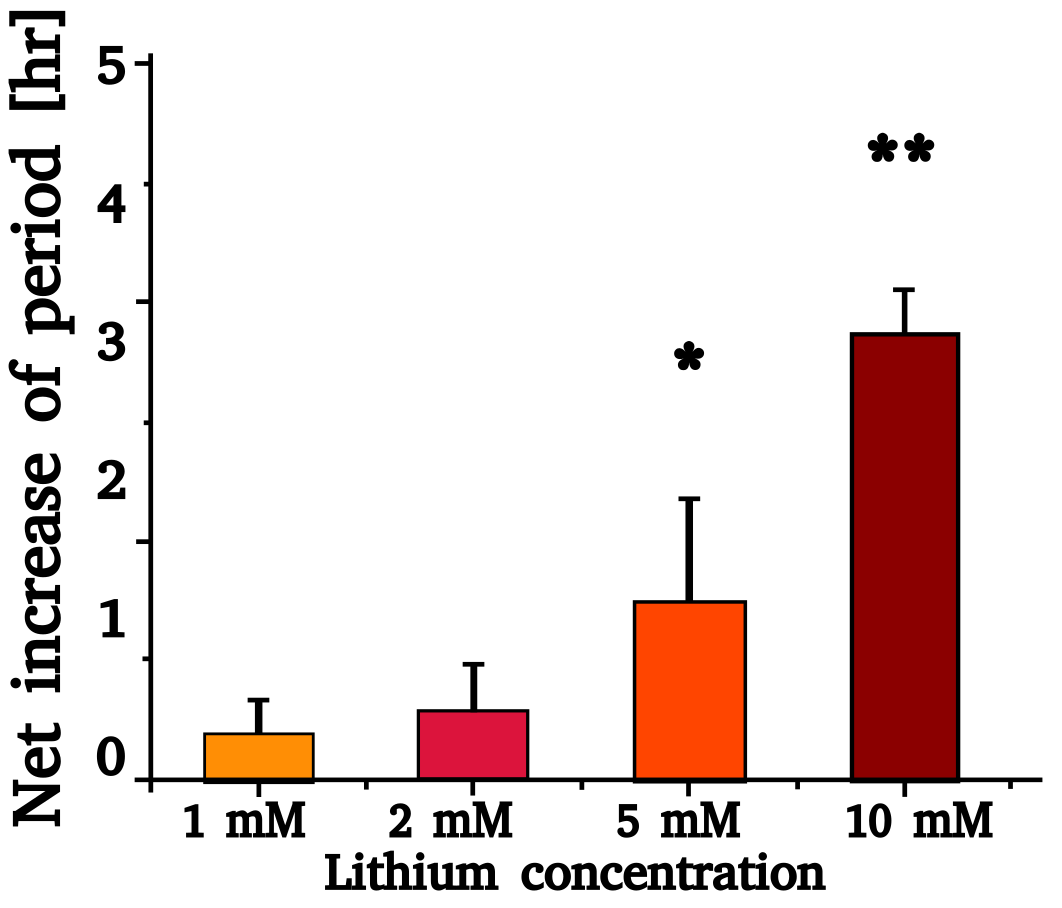}

	\end{minipage}

	\caption{Impact of lithium on circadian oscillations in PER2::LUC expression in mice lung slices \textit{in vitro}. The vertical blue line indicates addition of lithium salt. The visible transient behaviour is compensated for by prolonged measurement. * indicates $p<0.05$ and ** $p<0.01$, sample size $n = 5$. Adapted from \citet{li_lithium_2012}.}
	\label{fig:Li2012}
\end{figure}

	\noindent\textit{Lithium and circadian rhythm}\\
	Lithium ions are known to induce strong changes in the circadian rhythm of various single- and multicellular organisms when applied at levels similar to or within an order of magnitude of therapeutic serum levels of human patients. The circadian rhythm is measured through the fluctuations in metabolic markers and various forms of activity exhibited by the subject studied. The exact procedures needed to model therapeutic treatment are not fully established in single cell cultures, tissue samples, and animal models. Despite the variability of methods, it is generally found that it is possible to lengthen the period and increase the amplitude of the internal biological clock of living tissue by sustained exposure to lithium ions \cite{li_lithium_2012, yoshikawa_lithium_2016, rohr_impact_2022, mccarthy_genetic_2013, mishra_circadian_2021} (see figure \ref{fig:Li2012}). Lithium sensitivity can differ more than an order of magnitude between various cell and tissue types and species,  and likewise \textit{in vitro} versus \textit{in vivo} conditions can greatly influence the concentration needed to obtain these results. For example, lung tissue slices require higher concentrations than suprachiasmatic nucleus tissue slices \cite{li_lithium_2012}, and kidney cells are ten-fold more sensitive \textit{in vivo} than \textit{in vitro} (1mM vs 10mM) \cite{rohr_impact_2022, mccarthy_genetic_2013}. Even within a sample of the same species under the same conditions, considerable variability is observed and, in animal models, this may not correlate well with serum concentration \cite{rohr_impact_2022}. \\

The differential response observed in bipolar patients is reflected at a cellular level in human fibroblasts (a model for diurnal cellular rhythm), stem cell differentiated neural progenitor cells, and excitatory, glutamatergic neurons \cite{mccarthy_chronotype_2019, mishra_circadian_2021}. Across cell types there are similar, notable differences in the rhythm of metabolic markers pre- and post-treatment. Cells of lithium responders seems to adapt in manner similar, but weaker than control, while cells of non-responding patients react very little, if at all. Here 'lithium responder' is a much broader and less well-defined term that can include up to a third of the  patient population, and the sample sizes of these studies are still small due to the advanced techniques used. \\

\begin{figure}
	\centering
	\includegraphics[width=0.7\linewidth]{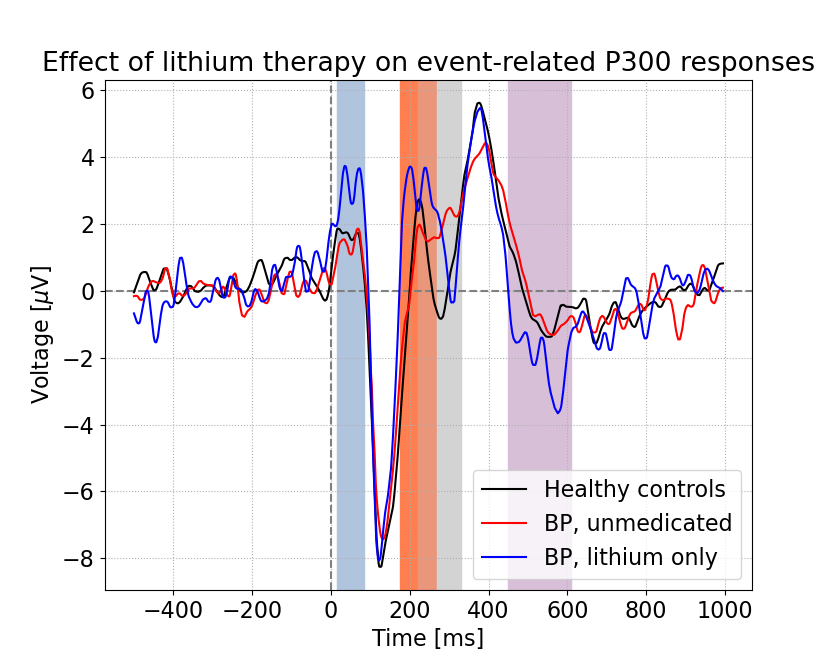}
	\caption{Impact of lithium on brain oscillations. Synchronized activity of large populations of neurons in the cerebral cortex is measured in bipolar patients free of medication, bipolar patients undergoing long-term lithium therapy and healthy controls. Each subject is exposed to a simple and repetitive sound pattern, occasionally disrupted by an 'unexpected' sound (auditory oddball experiments). The brain's response to this outlier is measured, and averaged within each group. Chronic lithium treatment noticeably enhances amplitudes in marked regions: blue, orange, light grey and purple. The impact of lithium therapy on healthy controls was not tested. Adapted from \citet{atagun_brain_2016}.}
	\label{fig:Atagun2016}
\end{figure}

Bipolar rhythms show strong dampening \cite{rohr_impact_2022} and lower amplitude \cite{mishra_circadian_2021, mccarthy_chronotype_2019} for both stem-cell differentiated neurons and fibroblasts. Though the non-responding subgroup performs notably worse than the lithium-responding subgroup, neither the dampening nor the amplitude are improved by long term lithium exposure and therefore not likely to be a target for lithium therapy \cite{mishra_circadian_2021}. In single-cell cultures, differentiated neurons from bipolar patients were desynchronized from control populations \cite{mishra_circadian_2021}. It is hypothesized that circadian rhythms in bipolar cell populations are less coordinated and have difficulity synchronizing to 'zeitgebers' \cite{mccarthy_genetic_2013}, since lithium exposure improves synchronization in response to change in the growth medium \cite{mccarthy_genetic_2013}. Lithium's ability to improve neural synchronization is also seen in electroencephalograms (EEGs) that measure the synchronized activity of large populations of neurons in the cerebral cortex \cite{atagun_brain_2016}. Medication-free bipolar patients show decreased amplitudes and reduced coherence in brain oscillations, while brain oscillations in lithium-treated patients are amplified compared to both medication-free subjects and control (see figure \ref{fig:Atagun2016}).\\

The clear differential effect is seen in the period elongation. Pre-treatment, the three subgroups in the neuron study had similar periods, while in fibroblasts, the period was notably longer in the bipolar subgroups, especially so for non-responders. Regardless of the differences between cell-types, chronic lithium treatment prolongs the circadian period of control and lithium-responding subpopulations but have no significant impact on the non-responding group \cite{mccarthy_genetic_2013, mishra_circadian_2021}.\\

Yet caution should be exercised. At higher doses (for fibroblasts: 10 mM - 20 mM, 2-7 days), the period-elongating effect of lithium response increases with lithium concentration \cite{mccarthy_genetic_2013, li_lithium_2012}. This does not seem to be the case specifically for the lithium-responding bipolar population at low doses (1mM, 5 days). Here, lithium treatment \textit{shortens} the period of lithium responder fibroblasts \cite{mccarthy_chronotype_2019}. Furthermore, \citet{mccarthy_genetic_2013} found increased amplitudes in control fibroblasts at an exposure of 1 mM (7 days) \cite{mccarthy_genetic_2013}, though fibroblasts generally do not respond at this concentration \cite{li_lithium_2012, mishra_circadian_2021}. This underlines the difficulty of characterizing lithium response.  Lithium response is likely a question of dose, duration and metabolic rate. The effective metabolic rate might not only depend on the inherent ability to produce a response, but also on the general metabolic rate of the organism, which is susceptible to other variables. One such variable is the mental state of the patient. In mania, the resting metabolism is elevated \cite{caliyurt_resting_2009}, and simultaneously acute mania can be treated effectively at doses considerably higher than what is recommended for chronic lithium therapy \cite{hsu_differences_2022}. To model the system correctly \textit{in vivo}, it is needed not only to monitor lithium concentration and duration of treatment, but also the level of metabolic activity in the sample.\\

Long-term lithium therapy can affect heart rhythms \cite{rosenqvist_sinus_1993, labbate_lithium_2023}, which is why most medical practitioners supplement patients' regular bloodwork with electrocardiograms (ECGs). The most common long-term effect observed is a slowing of heart rhythm (sinus bradycardia) \cite{mehta_lithiuminduced_2017}, which is usually benign. This is reversible by discontinuing lithium therapy \cite{manor_case_1983, ataallah_rare_2020}. It is possible that this effect is of a similar nature to the period elongation of the diurnal rhythm, but occurring at a different time scale. It is worthwhile to investigate whether cardiac tissue shows a similar differential response to lithium therapy. \\

	\noindent\textit{Lithium reverses neuronal hyperexcitability}\\
	Another interesting phenomena discovered through stem cell differentiation is lithium's effect on the excitability of nerves. Stem cell differentiated neurons derived from bipolar patients have been shown to be hyperexcitable and fire erratically compared to control, but this abnormality disappears in presence of therapeutic levels of lithium \cite{mertens_differential_2015, stern_neurons_2018, santos_deficient_2021}. This is puzzling, since acute lithium makes neurons hyperexcitable \cite{butler-munro_lithium_2010} and lithium overdoses can lead to epilepsy like firing and seizures \cite{izsak_differential_2021, yip_lithium_2007, schou_biology_1957}. This suggests that lithium response is the body's attempt to reverse the physiological impact of lithium. Whether lithium-responsive hyperexcitability exists in headaches is not known, but migraines have been associated with hyperactivity, specifically in cortical neurons \cite{scheffer_migraine_2013, aurora_transcranial_1998}. \\

Along with neuron hyperexcitability, \citet{mertens_differential_2015} also found decreased mitochondrial function in samples from bipolar patients \cite{malhi_therapeutic_2016,jakobsson_towards_2017, pereira_mitochondrial_2018}, associated with decreased mitochondrial membrane potential and size. These irregularities would be ameliorated in lithium-responders along with the hyperexcitability \cite{ mertens_differential_2015}. Brain tissue of deceased bipolar and schizophrenic patients show decreased pH levels, an indication that dysregulated brain metabolism causes mitochondria to shift to anaerobic respiration, thereby increasing lactate levels \cite{hagihara_decreased_2018, maurer_lithium-induced_2009, pereira_mitochondrial_2018}. It has been hypothesized that faulty ketosis during sleep is the cause of such an energy deficiency \cite{choi_sleep_2024}.
Furthermore, an acidic environment enhances neural excitability \cite{fillafer_excitation_2016}, potentially creating a feedback cycle that can tire out the body's adaptation, which might explain why lithium treatment is most efficient early in the course of illness \cite{kessing_starting_2014}. Counterintuitively, short term lithium exposure \textit{enhances} mitochondrial oxidative phosphorylation in human brain tissue \cite{maurer_lithium-induced_2009, pereira_mitochondrial_2018}, which would seem detrimental to a brain low on oxygen due to excessive activity such as in mania, lithium therapy's primary target. Again, the therapeutic effect of lithium administration is in contrast to its acute impact on the body.\\

It seems that lithium impacts the state of the cell and body through a general mechanism, but not all bodies are able to adapt. Certain target areas are more interesting than others, such as cellular clocks, nerve reactivity, mitochondrial function, and possibly membrane susceptibility to viral attack. Here, the consequences of adaptation leads to visible curative power. Neurons and mitochondria have in common that their primary functions are very much tied to their membrane function, so it is reasonable to assume that the metabolic response has to do with global membrane adaptation. Lithium response might broadly affect some characteristic membrane quality, that is targeted or indirectly affected by pathologies.\\

	\noindent\textit{Membrane lipid adaptation}\\
	When rats are subjected to chronic lithium exposure, their synaptic cell membranes undergo lipid compositional changes, which was measured by \citet{joseph_systemic_1987} and \citet{lopez-corcuera_change_1988} in the late 80s, and \citet{fisar_effect_2005} and \citet{pettegrew_effects_2001} in the early 2000s. In the first 3 studies, rats were treated with lithium for 4 weeks, while in the latter the period was 2 weeks. The phospholipid species of interest were the zwitterions, phosphatidylethanolamine (PE) and phosphatidylcholine (PC), and the anions, phosphatidylserine (PS) and phosphatidylinositol (PI), the latter two being important regulators of neural function \cite{li_review_2015}. Though all studies report changes in phospholipid expression, it is not possible to provide a detailed account of the membrane adjustment in phospholipid species. Generally the sample sizes were small, and in the cases of \citet{joseph_systemic_1987} and \citet{fisar_effect_2005} there was a lack of statistical significance. However, \citet{lopez-corcuera_change_1988} found a considerable change in zwitterionic lipid expression with a 30\% increase in PE levels and a 28\% decrease in PC levels. Both \citet{lopez-corcuera_change_1988} and \citet{pettegrew_effects_2001} found small decreases in anionic lipids of 7\% and 3\% in PS, respectively, and \citet{pettegrew_effects_2001} additionally found a a 8\% decrease in PI, while no noticeable change was reported by \citet{lopez-corcuera_change_1988}. \\

Additionally, \citet{lopez-corcuera_change_1988} and \citet{fisar_effect_2005} measured cholesterol.  With cholesterol, there seems to be a clearer picture: \citet{fisar_effect_2005} and \citet{lopez-corcuera_change_1988} found significant increases in membrane cholesterol concentration, in total and relative to membrane protein, respectively. Though \citet{lopez-corcuera_change_1988} found no significant change in cholesterol-to-phospholipid ratio, \citet{fisar_effect_2005} found the CL/PL ratio increased by 16 \%. Since manic patients have abnormally low cholesterol levels \cite{fusar-poli_lipid_2020}, this was a first indication of the therapeutic nature of lipid response. \\

In human patients, lithium therapy is known to elevate lipid serum levels, such as cholesterol and triglycerides, both in studies and clinical practice \cite{aliyazicioglu_treatment_2007, joshi_evidence_2019, sussulini_metabolic_2009, ranjekar_decreased_2003}. This is also indirectly evident through lithium therapy's heavy toll on the thyroid gland. Up to 32\% of patients in long-term lithium therapy end up developing a thyroid disorder, primarily a distinct version of hypothydorism \cite{joseph_long-term_2023, kraszewska_thyroid_2019, kraszewska_no_2019}, which is associated with elevated cholesterol and triglycerides levels \cite{mavromati_hypothyroidism-associated_2021}. Interestingly,  sinus bradycardia is also common in hypothydorism.\\

At a grander scale, the major genetic pathways associated with lithium response in bipolar disorder are likewise concerned with lipid metabolism.\\

Lithium response has a general, restorative effect on neural function \cite{lyoo_lithium-induced_2010, malhi_therapeutic_2016}, which has been correlated with changes in a myriad of biomarkers, most notably the proteins GSK-3 and BDNF (In the latter, differential response has been documented \cite{m_bdnf_2008, suwalska_serum_2010}).  Attention has been paid to inositol lipid metabolism and the expression of the enzyme myo-inositol monophosphatase, which can be significantly altered in patients with bipolar disorder or by general lithium administration \cite{ sherman_evidence_1981, belmaker_behavioral_1998}. Phosphatidylinositols makes up around 10\% of lipids in eukaryote cells \cite{rattay_viruses_2023} and are highly prevalent in myelin \cite{irvine_short_2016}. Lithium therapy reduces the expression of phosphatidylinositol, which is associated with an upregulation in phospholipase activity (correllated with cellular lithium response \cite{kimata_synaptic_2012}) and  downregulation of the inositol-producing IMPase. This change in membrane composition has been associated with disrupted organization of polar compontents at synaptic ends \cite{tanizawa_inositol_2006, kimata_synaptic_2012} and improved synaptic function \cite{paquette_human_2023, kim_lithium_2009}, establishing yet another link between membrane lipid composition and excitability. For these reasons, inositol depletion has previously been proposed as a primary mechanism of lithium therapy, but found not to match the onset of treatment response and therefore believed not to be the primary mode of action \cite{malhi_therapeutic_2016, toker_lithium_2014, moore_temporal_1999}. \\

In the genome-wide association studies by \citet{ikeda_genome-wide_2018} and \citet{song_genome-wide_2016} both bipolar disorder and lithium response have been linked to genes involved in lipid metabolism. \citet{ikeda_genome-wide_2018} found bipolar illness to be associated with a gene in the fatty acid desaturase family (FADS), associated with regulation of plasma blood lipids and omega-3 and -6 polyunsaturated fatty acids. When comparing lithium responders with healthy controls, \citet{song_genome-wide_2016} were able to isolate the gene, SESTD1, related to phospholipid regulation. While \citet{song_genome-wide_2016} found no significant genetic profile for lithium responsiveness within the bipolar population, other studies have reproducibly correlated multiple genes with lithium response within the bipolar population, associated with cellular processes such as calcium signaling, inflammation response and neuronal function \cite{hou_genetic_2016, le_clerc_hla-drb1_2021, hou_genome-wide_2016}. The growing number of sites associated with lithium response suggests that there might be myriad contributions to a decreased response to lithium. The study by \citet{song_genome-wide_2016} is then especially interesting, since it seems to suggest that specifically within the group of lithium-responding bipolar patients, a few genetic variations are hindering them from responding to the same extent as healthy controls.\\

Since this review focuses on the thermodynamics of the cell membrane, we will not give a detailed account of exactly how the organism facilitates lithium response by changing metabolic pathways through gene expression. We direct the reader to reviews written by \citet{malhi_therapeutic_2016, gao_chapter_2021, can_molecular_2014} and \citet{jakobsson_towards_2017}. \\

	\noindent\textit{Connection to oxidative stress and inflammation}\\
	How does this supposed lithium-induced adjustment in membrane lipid content then tie to lithium response in mood disorders and headaches? Firstly, lipid metabolism is involved in some of the key physiological aspects of both disorders, namely cellular stress and inflammation.\\

Cellular activity produces harmful remnants from respiration, known as reactive oxygen species (ROS), that accumulate in the cellular environment if these are not properly counteracted by the body. This accumulation exerts an 'oxidative stress' on cellular function by oxidation of proteins, lipids, DNA, and sugars \cite{gross_mitochondrial_2021}. Nerve tissue is especially  susceptible to oxidation, due to its high concentration of iron \cite{gerlach_altered_1994} and polyunsaturated fatty acids \cite{borst_oxidation_2000}. Excessive nerve activity can thus destroy nerve and mitochondrial function (excitoxicity) \cite{wang_energy_2023}. Brain tissue, that operates at a very high oxygen consumption, is therefore vulnerable to unregulated nerve activity \cite{aliyazicioglu_treatment_2007}.\\

A large presence of oxidative species in nerve tissue can be caused by overly active nerves or bad regulation of chemical species that activate the nerve. Two classes of such agents are the known 'stress hormones' such as cortisol, adrenaline, and noradrenaline, and 'excitatory neurotransmitters' such as glutamate and dopamine. Bipolar patients seem to get the worst of both worlds; their neurons are overly active \cite{mertens_differential_2015} \textit{and} mood disorders are associated with irregularities in dopamine levels and HPA-axis regulation \cite{gray_lithiums_2013, joshi_evidence_2019, palsson_altered_2023}, elevated levels of lipid peroxidation and antioxidant response \cite{machado-vieira_oxidative_2007, kuloglu_lipid_2002}, disrupted cellular calcium-reactivity, and excessive response to corticostroids \cite{mccarthy_calcium_2016, gray_lithiums_2013, kenna_psychiatric_2011, finley_drug_2016}. As mentioned earlier, chronic lithium treatment has the ability to reverse the hyperexcited and erratic behaviour in samples from responding patients, and it is found that while acute lithium activates the nerve \cite{mccarthy_calcium_2016, izsak_differential_2021}, in general, chronic lithium decreases nerve and mitochondrial sensitivity to excitatory agents in human \cite{mccarthy_calcium_2016} and animal models \cite{chen_inhibition_1996, shalbuyeva_lithium_2007, schou_biology_1957, drummond_lithium_1987} (A 'serotogernic effect' \cite{malhi_therapeutic_2016} \cite{malhi_potential_2013}). Although lithium has been shown to induce changes in bipolar patients' HPA axis regulation, some irregularities do persist in stabilized individuals, which could give a clue to underlying pathology \cite{gray_lithiums_2013}. \\

Lithium's ability to decrease cellular stress \cite{malhi_therapeutic_2016, gray_lithiums_2013} is evident in both humans \cite{machado-vieira_oxidative_2007, aliyazicioglu_treatment_2007, pisanu_differences_2020} and animals \cite{vosahlikova_therapeutic_2021, shao_chronic_2005, carli_effects_1997} and can therefore be assumed to be a consequence of the systemic adaptation. It fits well that serum lipid peroxides have been measured to decrease, while lipid serum levels rise in bipolar patients undergoing lithium treatment \cite{aliyazicioglu_treatment_2007}. \\

	\noindent\textit{Connection to calcium reactivity}\\
	Intracellular calcium movement is a key 'second messenger' in excitatory tissue. Calcium buffers transmit the presence of excitatory agents at the membrane interface to the cell interior through a cascade effect, starting by an arrival of calcium ions at the outer layer of the cell membrane that instigates release of calcium from intracellular buffering membranes and macromolecules, such as the endoplasmatic reticulum, mitochondria, and mitochondria-associated membranes (MAMs) \cite{berridge_inositol_1987, britti_mitochondrial_2018}.  Several mechanisms previously mentioned are directly involved in or dependent on intracellular calcium release, such as: mitochondrial homeostasis and energy production \cite{walkon_calcium_2022}, circadian rhythm \cite{mccarthy_calcium_2016}, and inositol metabolism \cite{berridge_inositol_1987, gray_lithiums_2013, drummond_lithium_1987}. \\ 

\begin{figure}
	\centering
	\includegraphics[width=0.7\linewidth]{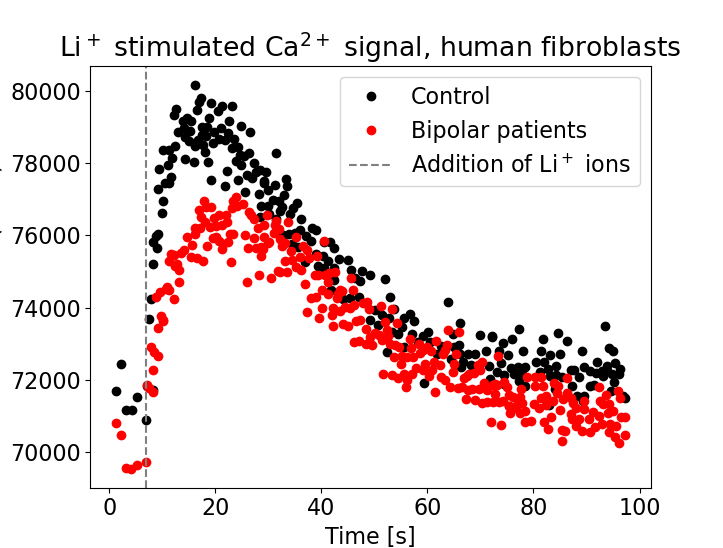}
	\caption{Release of intracellular calcium in human fibroblasts after acute lithium administration. The grey, broken line indicates the addition of lithium ions. In acute form, lithium ions have an activating effect on the cell, releasing membrane-associated calcium from interior membranes. Fibroblasts from bipolar patients do not produce as strong a response as control, an indication that bipolar illness affects membrane buffer properties. Adapted from \citet{mccarthy_calcium_2016}.}
	\label{fig:McCarthy2016}
\end{figure}

Bipolar patients show significantly altered cellular calcium reactivity.  \citet{mertens_differential_2015} found an increased frequency of intracellular calcium transients in neuron populations differentiated from bipolar fibroblasts, and \citet{mccarthy_calcium_2016} have found the calcium release in bipolar fibroblasts to be less pronounced compared to control (See figure \ref{fig:McCarthy2016}). The combination of the two studies suggest that the cellular buffer becomes less responsive during excessive neuronal stimulation. Additionally, \citet{mertens_differential_2015} found that long-term lithium treatment reduces the frequency of calcium transients in lithium-responders only, while control and non-responders were unperturbed. Lithium is an excitatory agent, inducing intracellular calcium release in excitatory tissue, while repeated or chronic exposure depletes this response over time \cite{chen_inhibition_1996, drummond_lithium_1987, shalbuyeva_lithium_2007}. Lithium-response thus further desensitizes the cell. Why the control population does not show similar changes in calcium reactivity when subjected to lithium treatment can be attributed to the non-linear behaviour of membrane buffer properties, which I will come back to in the section \textit{Membrane buffers}.\\

	\noindent\textit{A fundamental mechanism}\\
	A picture starts to emerge: Long-term lithium exposure numbs the nerves to the presence of excitatory agents, naturally present in bipolar patients due to a faulty stress response. Initially, lithium makes the situation \textit{worse}, stimulating calcium release \cite{mccarthy_calcium_2016}, increasing hyperexcitability \cite{izsak_differential_2021}, enhancing mitochondrial oxidative phosphorylation \cite{maurer_lithium-induced_2009, pereira_mitochondrial_2018}, and possibly even inducing apoptosis in younger neurons \cite{dmello_lithium_1994}. Therapy occurs when the body counteracts this effect by changing the membrane composition \cite{vosahlikova_therapeutic_2021}. The nerves are no longer erratic and overactive, produce less of the inflammatory agents, and the body heals. But how does a change of lipid composition in the membrane affect the reactivity of the cell?\\

Changes in lipid composition alter the physical properties of the lipid membrane. In the case of lithium response, significant changes in membrane fluidity were observed by \citet{herrero_chronic_1987} and \citet{lopez-corcuera_change_1988}. In their study of rats treated with lithium long-term, \citet{lopez-corcuera_change_1988} found the changes in synaptic lipid species to be accompanied by an increase in the phospholipid-to-protein ratio of 20\%, a general decrease in hydrocarbon chain saturation, and increased membrane fluidity measured through fluorescence spectroscopy. As we shall see in the next chapter, lithium has a unique ability to modulate the electrostatic potential and lipid phase in both pure lipid bilayer vesicles and biological membranes, unlike ions of similar ionic radius and charge. We will revisit an old hypothesis by \citet{hauser_crystallization_1981} \cite{hauser_interactions_1983} that lithium interacts with the body primarily through lipids, altering crucial properties of cellular constituents, such as buffer ability, morphology, and ability to conduct signals, which induces a strong metabolic response in the healthy organism. \\

\section{Lithium on a molecular scale}

	The backbone of cellular membranes are polar phospholipid bilayers, where proteins, sugars, and other lipid species are inserted or attached to \cite{lee_lipid_1977}.
The lipids can be packed in various ways, since the conformation of their backbones respond to their surroundings. This can be used to manipulate the lipid bilayer into distinct, structural states impacting overall membrane  morphology and function. In pure lipid membranes, the lipids change conformation in such a coordinated manner that the change of state appears like a first-order phase transition. Biological lipid membranes change phase between a fluid crystal and a more ordered, gel-like state. Ions influence the lipid configuration in a manner similar to proteins, by interacting with the polar surface and changing lipid hydration \cite{cevc_how_1987, cevc_mechanism_1986, binder_effect_2002}, but membrane state can be affected by many other variables, such as temperature and pressure \cite{binder_effect_2002, jacobson_phase_1975}. \\

Ions interact with membranes in many ways. They can alter the structure of the water in contact with the membrane, change the conformation of membrane lipids and proteins inserted or attached to the membrane, and influence the interaction between membranes \cite{cunningham_influence_1986}. We will mainly investigate lithium's effect on the lipid membrane in comparison to other alkali ions and calcium.\\ 

Small alkali ions, such as the naturally prevalent sodium and potassium, associate only weakly with binding sites in the headgroup region of the lipid bilayer. They can have a lot of water associated to them, and their presence can both increase and decrease hydration of phospholipid membranes, depending on what is entropically favorable for the system \cite{binder_effect_2002, klasczyk_interactions_2010, collins_charge_1997}. Calcium, on the other hand, binds strongly to the headgroups of the lipid bilayer and liberates them from their associated water, partially dehydrating the membrane \cite{binder_effect_2002}. This mechanism is likewise entropically driven \cite{sinn_binding_2006}. \\

Interestingly, lithium has a uniquely high membrane affinity in both pure-lipid membranes \cite{jakobsson_towards_2017, kotynska_association_2017, klasczyk_interactions_2010}  and biological membranes (15-18 times higher than sodium) \cite{ehrlich_lithium_1980}. The lithium ion also has an extraordinary dehydrating ability \cite{hauser_interactions_1983, casal_infrared_1987} compared to other alkali ions. On the surface, it seems to resemble the divalent calcium ion more than its fellow alkali ions in its interactions with the phospholipid membrane \cite{hauser_interactions_1983, casal_infrared_1987}.\\

There are mainly two types of events that can take place when ions influence membrane phase. Ions with low membrane affinity screen the surface charges through ionic strength \cite{trauble_electrostatic_1974}, while ions of high membrane affinity adsorp in the lipid bilayer \cite{trauble_electrostatic_1974} by associating or binding to lipid headgroups. In the following sections, we will investigate how well lithium does either, in comparison with the biologically prevalent ions sodium, potassium, and calcium.\\


	\noindent\textit{Electrostatic screening of membrane potential}\\
	When ions of low membrane affinity, such as small metal cations, distribute themselves in the water surrounding a polar membrane, they interact with the membrane primarily through their electrical field, and have only a small effect on hydration and lipid structure \cite{binder_effect_2002}.  Alkali ions 'screen' the negative charges of lipid headgroups without strong association. This can decrease or modify the electrostatic potential across the bilayer.\\

Lithium is effective at screening charged lipid bilayers. A molecular dynamics study by \citet{lopez_cascales_effect_1997} investigated lithium and sodium ions' screening ability in an anionic phospatidylserine bilayer (DPPS) and found that lithium, unlike sodium, was very effective at decreasing the transmembrane potential, which aligns well with lithium's high membrane affinity. This can be attributed to lithium's high surface charge density in solution, which is about 5 times higher than that of sodium \cite{rayner-canham_descriptive_2010}. \\

Behaviour in single ion assays is not entirely translatable to the cellular environment, though, since ion species interact and can compete for access to the cell membrane. One needs to determine how lithium interacts with other ions, such as sodium, potassium and calcium, at the lipid bilayer interface. Here, \citeauthor{lopez_cascales_effect_1997} created a simulation of mixed lithium and sodium at equal doses. Sodium counteracts lithium membrane association and lithium screens less effectively, while sodium itself has been fully displaced into the surrounding water. This phenomena has also been observed \textit{in vitro} red blood cells and kidney cortex slices, where lithium displaces both sodium and potassium \cite{schou_biology_1957}, and exemplifies how electrolyte balance might drastically alter and be altered by lithium treatment. Considerable electrolyte imbalances have been documented in affective episodes and patients undergoing lithium therapy \cite{coppen_mineral_1966, mifsud_lithium-associated_2020, nechifor_magnesium_2011}. Likewise lithium ions have been documented to compete with the divalent cations calcium and magnesium for anionic lipid headgroups \cite{fossel_lithium-7_1985}. \\


\noindent\textit{Lipid phase response to ion screening}\\
Ionic screening increases the fluidity of the lipid bilayer, thereby lowering the temperature at which the membrane switches from gel to fluid phase, i.e. \textit{melting point depression}. When screening of headgroup charges reduces surface potential of the membrane, the electrostatic repulsion between headgroups lessen, and lipids can move more freely and closer to each other \cite{overath_phase_1973, trauble_electrostatic_1974}. This has been observed at low doses of sodium and potassium (and lithium, occasionally) in both anionic and zwitterionic bilayer vesicles, such as phosphatidic acid (DHPA) \cite{trauble_electrostatic_1974} and phosphatidylcholine (DPPC) \cite{eklund_monovalent_1989, binder_effect_2002}. Since lithium is so effective at screening the membrane potential, one would expect it to likewise induce melting point depression in lipid bilayers. This is rarely the case, however, due to interactions with lipid head groups. \\

	\noindent\textit{Membrane adsorption}\\
	Strength of the association between a molecular binding site and metal ion is overall determined by charge \cite{kotynska_association_2017}. The monovalent cations, sodium and potassium have minuscule membrane affinities in comparison to the divalent calcium. Calcium binds strongly to lipid headgroups with full or partial charge neutralization \cite{hauser_interactions_1976, trauble_electrostatic_1974, shih_membrane_2018}, and can even coordinate with multiple lipids at the same time  \cite{binder_effect_2002}. \\

Both lipids and proteins show high specificity to ion species \cite{hauser_interactions_1976, binder_effect_2002}. The variety of binding sites and geometry of the macromolecule amplifies the small differences in ionic radius and mass of metal cations, such that two ions of the same charge and similar ionic radius and mass might associate itself very differently to the same molecule. \\

Among ions of the same charge, there exists a hierarchy in membrane affinity dictated by differences in surface charge density. This fine structure, or lyotropic series, is similar to the Hoffmeister series of protein precipitation \cite{hauser_interactions_1976, kotynska_association_2017, klasczyk_interactions_2010, cunningham_influence_1986}. Compared to other alkali ions, lithium has a very high surface charge, and this is the cause of lithium's unique membrane affinity \cite{kotynska_association_2017}. Lithium ions adsorp into the membrane surface, associating with membrane headgroups even after effective screening \cite{cevc_thermodynamic_1985} and sometimes over-compensating for the lipid charge \cite{kotynska_association_2017, klasczyk_interactions_2010}. Lithium adsorption resembles proton adsorption -- both ions associate in a similar manner to phosphate and carbonyl/carboxylate groups in the lipid headgroup region \cite{jakobsson_towards_2017, cunningham_influence_1986, cevc_thermodynamic_1985}. \\

\noindent\textit{Lipid phase response to ion binding}\\
When ions associate to lipid headgroups, they induce conformational changes in the glycerol backbone of the lipid that greatly affect the liquid-crystalline phase behaviour of phospholipid bilayers \cite{binder_effect_2002, casal_infrared_1987, cevc_thermodynamic_1985}. The conformational changes decrease lipid interaction with water by sharing hydration shells or even dispelling the water from the binding sites of lipid headgroups \cite{binder_effect_2002}.\\

The dehydration increases lipid interaction, the packing of hydrocarbon chains becomes more ordered, and the melting temperature rises \cite{binder_effect_2002,hauser_interactions_1976}. The phase-altering effects of di- \cite{trauble_electrostatic_1974, binder_effect_2002}, tri- \cite{bhalla_effects_2009, silva_aluminum_2002} and quadrivalent cations \cite{ahmadvand_changes_2010} are already well-documented in phospholipid and biological membranes, while lithium phase modulation is relatively unknown. Lithium has a surface charge density higher than any other alkali ion \cite{rayner-canham_descriptive_2010}, and similar to that of calcium, but lithium membrane affinity and phase modulation are still orders of magnitude weaker than calcium \cite{cevc_thermodynamic_1985, hauser_interactions_1983, jacobson_phase_1975}. Lithium also has a much smaller ionic radius, enabling it to pierce deeper into the lipid headgroup region. This causes the lipids in the lithium stabilized gel state to exhibit a characteristic tilt more similar to proton binding \cite{cevc_thermodynamic_1985, casal_infrared_1987}. After a more in depth look, the lyotropic behaviour of lithium ions turns out to be rather unique, differing from both divalent calcium and other alkali ions.\\

\begin{figure}
	\centering
	\includegraphics[width=0.9\linewidth]{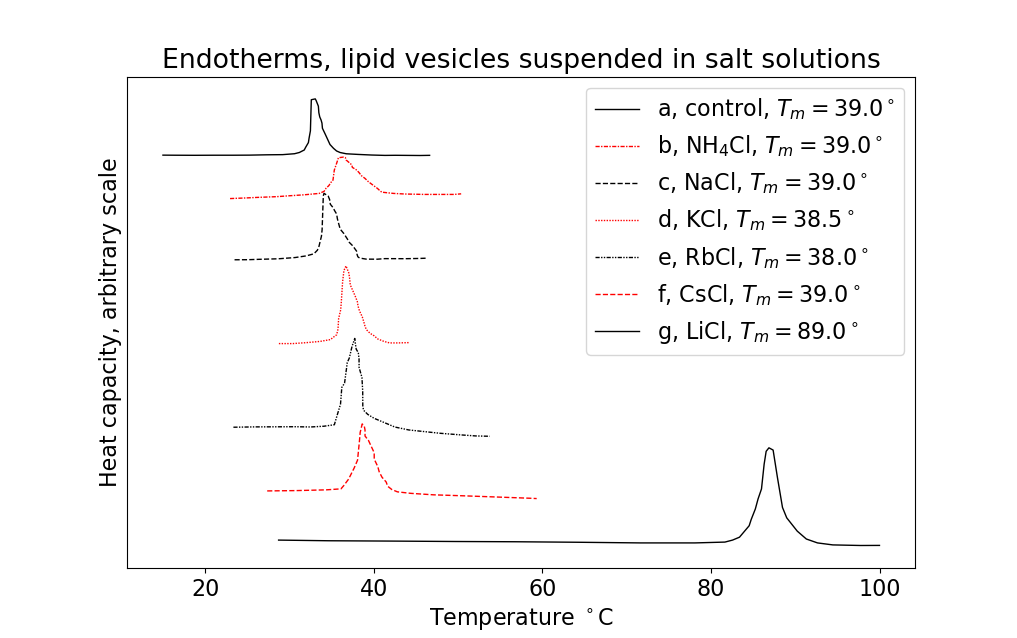}
	\caption{Melting profiles of DMPS bilayer vesicles in monovalent ion dispersions, 0.5 M (pH 5.8-6.3). In the transition regime between the pure fluid and gel phase, the membrane lipids enjoy increased mobility and other degrees of freedom, which causes a transient peak in heat capacity. Data adapted from \citet{hauser_interactions_1983}. }
	\label{fig:Hauser1983}
\end{figure}

	\noindent\textit{Experiments in lipid vesicle suspensions}\\
\citet{hauser_crystallization_1981}\cite{hauser_interactions_1983} showed in the 1980's that it is possible to shift the melting point of phosphatidylserine bilayer vesicles (DPPS, DMPS, DLPS, DSPS) by 40-50$^{\circ}$C in a dispersion of 0.5 M LiCl (pH 5.8-6.3, see figure \ref{fig:Hauser1983}). In comparison, 0.5M NaCl or KCl had no visible impact or even lowered the main transition by 1-2 degrees. At lower doses of lithium ($<<$0.5 M), or in zwitterionic lipid bilayers (such as DPPC and DPPG), the shift is an order of magnitude smaller, if present at all. It is still evident that lithium has a unique influence on the melting profiles compared to other alkali ions \cite{hauser_crystallization_1981, cunningham_influence_1986, eklund_monovalent_1989, kamp_numerical_2021, binder_effect_2002}. \\

Generally, ion headgroup binding impacts the lipid membrane much more than ion-induced changes in the water at the membrane interface \cite{cunningham_influence_1986}. But lithium headgroup association might not alone explain the excessive dehydration taking place at high molarities in anionic lipids. At sufficiently high ion concentrations (Lithium and sodium: $>$0.5 M and $>$0.6 M for DPPG at pH 7.4 and DPPS at pH 6.5 - 7 respectively \cite{cevc_thermodynamic_1985} \cite{eklund_monovalent_1989}, see also \cite{jacobson_phase_1975}), a third mechanism is dominant, caused by alkali ions' hygroscopic abilities. Small metallic cations hydrate strongly, and in sufficiently high concentration they steal water from intermembrane space while screening the intermembrane potential, which increases lipid-lipid interaction, and thereby increases the melting point of the bilayer \cite{hauser_crystallization_1981, hauser_interactions_1983, hauser_aspects_1984, jaehnig_electrostatic_1979, eklund_monovalent_1989}. Since anionic lipids can accumulate large amounts of intermembrane water \cite{hauser_aspects_1984}, and lithium has excellent screening abilities, this effect likely gets exponentiated.\\

\noindent\textit{Reversibility of lipid phase transitions and the importance of lipid charge}\\
Lipid conformational changes are, on local scale, reversible, but this does not mean, however, that large scale lipid dispersions will return to the exact same state passing in and out of transition, if the initial state of the membrane is slightly dehydrated  \cite{cevc_thermodynamic_1985, binder_effect_2002, chapman_lipid_1977, melchior_thermotropic_1976}. Configurational changes taking place in the lipid during phase transition can change the availability of charges. Formerly dehydrated headgroup moieties hidden deeper in the headgroup region get exposed to water, and they may not dehydrate when the lipid configuration changes back. The large shift of 40-50 degrees reported by \citet{hauser_interactions_1983} decreased by $\approx 20$ degrees after repeated heatings of the lipid dispersion, indicating the vesicles initally existed in a extraordinarily dehydrated state. It is not uncommon to see such a large hysteresis in data due to this phenomena \cite{eklund_monovalent_1989}. A similar hysteresis is seen by changing pH. An increase in pH from 7 to 9 in DPPA membranes doubles the elementary charges per polar group, lowering the melting temperature by 20 degrees \cite{trauble_electrostatic_1974}.\\

As shown, negatively charged lipids display a much stronger reaction to the presence of lithium than neutral ones \cite{hauser_interactions_1976}. This is not primarily due to differences in headgroup interactions, but a question of recruitment to membrane surface in highly diluted systems \cite{binder_effect_2002}. Charged lipid bilayers have ion affinities at least 1000 times stronger than neutral lipids \cite{hauser_interactions_1976}. Additionally, ion affinities for the lipid membrane in aqueous solutions are highly intensified by reducing hydration \cite{binder_effect_2002}.\\

All effects mentioned can have considerable consequences for biological systems, where up to 20\% percent of membrane lipids are ionizable \cite{jaehnig_electrostatic_1979}, and bulk water is rare. Anionic lipids, such as phosphotidylserine and the doubly charged cardiolipin are especially prevalent in the peripheral and central nervous system \cite{hauser_interactions_1983} and mitochondria \cite{horvath_lipids_2013} respectively, where they can constitute up to 20\% of the lipid matrix \cite{jaehnig_electrostatic_1979}. It is thus entirely possible that significant lipid phase modulation can take place at lithium concentrations far below $\approx$1M.\\

\section{Relevance to biology}

	\noindent\textit{Biology at transition point}\\
	Many phenomena suggest that lipid phase contributes to cellular function. Biological membranes are close to lipid phase transition at physiological conditions, and studies in cellular adaptation indicate that maintaining this proximity is important for survival \cite{melchior_thermotropic_1976, overath_correlation_1970, muzic_melting_2019}. \\

Transition temperature, transition range, and influence of ionic strength has been found to be essentially the same for phospholipid bilayer vesicles, biological membranes, and whole cells \cite{overath_phase_1973}. This indicates that pure phospholipid membranes are good models for biological membranes under similar lipid composition and thermodynamic conditions. Bulk lyotropic behaviour of the earlier section is therefore applicable at cellular level, such as phase-dependent buffer ability, morphological changes, and the sensitivity to hydration level and ion concentration in the cellular environment. There is evidence that ionic influence on the activity level of muscles, excitable tissue, cilia, and spermatozoa follows the Hofmeister series \cite{schou_biology_1957}. \\

	\noindent\textit{Layers of complexity}\\
	Live biological membranes exhibit a diverse and constantly adaptive expression of membrane constituents and morphologies. Their transition range is broad due to the varied lengths and levels saturation of hydrocarbon tails  \cite{overath_phase_1973} and presence of cholesterol and membrane proteins \cite{overath_phase_1973, muzic_melting_2019, heimburg_lipid_2010, halstenberg_cholesterol-induced_1998}. Biological membranes often undergo several distinct stages due to presence of different lipid species or changes in membrane geometry \cite{muzic_melting_2019, schneider_network_1999}. \\

Lipids can form into domains depending on structure and conformation,
which can affect membrane morphology and create interfacial effects. The lipids residing at the domain interface are less coordinated with their surrounding lipids, and therefore free to fluctuate more, increasing the membrane permeability at the boundary \cite{heimburg_lipid_2010}. Large membrane proteins can induce phase boundaries in the surrounding lipid matrix, likewise creating pores \cite{heimburg_lipid_2010}. Furthermore, there can be an interplay of all mechanisms mentioned: Proteins might prefer one lipid phase above the other and accumulate in a lipid domain, thus amplifying the disruption of lipid cooperativity \cite{thilo_mechanistic_1977, melchior_thermotropic_1976, hauser_interactions_1976, hauser_interactions_1983, overath_phase_1973, papahadjopoulos_phase_1973, binder_effect_2002}. \\

And while some membrane constituents serve to modulate phase response, other components might deprive the membrane of this response all together. Cholesterol is an important regulator of membrane phase in biological systems \cite{bhalla_effects_2009}. Cholesterol functions as a plasticizer, decreasing phospholipid cooperativity and broadening the lipid phase transition \cite{farber_broad_2022, schrader_compressibility_2002}. At high concentrations, it abolishes the lipid phase transition and 'locks' the membrane in a semi-fluid state \cite{melchior_thermotropic_1976, papahadjopoulos_phase_1973}. One example of a tissue high in cholesterol and void of lipid phase transition is myelin \cite{schrader_compressibility_2002, melchior_thermotropic_1976}. As mentioned previously, lithium response is partially mediated through an increase in cholesterol content. Interestingly, \citet{fisar_effect_2005} reports that the measured increase in membrane cholesterol content in rats due to long-term lithium treatment does not entail an increase in fluidity measured by fluorescence anisotropy, indeed indicating that cholesterol counteracts the effect of lithium.\\

    \noindent\textit{Lipid phase as a fundamental mechanism}\\
	A good reason for staying near membrane lipid transition is the ease at which one can induce physical changes in the membrane. A small change in state variables, such as pH or temperature, can cause a disproportional change in the physical properties of the system. Manipulation of lipid phase is thus a fundamental and powerful tool that might be utilized by the cell in multiple membrane-mediated functions. \\

	\noindent\textit{Morphology}\\ 
When ions interact with lipid bilayers, they alter the surface pressure acting within the polar interface \cite{binder_effect_2002, liu_asymmetrical_2022}. This is especially the case when the presence of ions causes conformational changes in the backbone of phospholipids. When phospholipid membranes melt, they gain many degrees of freedom: their  hydrocarbon chains are free to rotate, the lipids diffuse laterally within the membrane \cite{trauble_electrostatic_1974} and even flip between bilayers. During transition, the lipid bilayer undergoes a peak in compressibility \cite{heimburg_mechanical_1998, schrader_compressibility_2002} proportional to the peak in heat capacity \cite{heimburg_thermal_2007}(chap. 4.9 \& 4.10.1) \& \cite{halstenberg_cholesterol-induced_1998, mosgaard_fluctuations_2013} and the lipid bilayer undergoes considerable physical changes. \\

In lesser degree, conformational changes in the backbone of phospholipids creates a phase separation in pure lipid membranes \cite{rappolt_salt-induced_2001} and induces domain formation in mixed lipid membranes \cite{binder_effect_2002}. When the lipid matrix reaches phase transition, the collective changes in lipid conformation cause the lipid bilayer to undergo large changes in area, thickness, and volume \cite{trauble_electrostatic_1974, heimburg_soliton_2005, heimburg_mechanical_1998} \cite{heimburg_thermal_2007}[chapt. 18.3.2.2]. As an example, pure DPPC bilayer vesicles undergo a 25 \% increase in area, a 4.7\% increase in volume, and a 16 \% decrease in thickness in the transition from gel to fluid \cite{heimburg_soliton_2005} \cite{heimburg_thermal_2007}[chapt. 18.3.2.2]. \\

These dimensional changes enable membrane curvature, vesicle formation, lipid vesicle aggregation, membrane fusion, and the formation of membrane networks \cite{hauser_interactions_1976, schneider_network_1999,  epand_effect_1986, binder_effect_2002}. By ion concentration, pressure, or other variables can the body thus realize large physical changes with little effort. This has many possible applications in biology, and was proposed in the 70s by researchers such as Overath, Träuble, and Hauser \cite{overath_phase_1973, trauble_electrostatic_1974, overath_lipid_1976, hauser_interactions_1976} to be the primary controlling mechanism behind cellular functions such as pinocytosis, release of neurotransmitters and hormones from vesicles, cell division, and nerve signaling. The increase membrane fluidity associated with lithium response will have an impact on morphology, and that may be the reason why long-term lithium therapy ameliorates abnormal mitochondrial shape and size in bipolar disorder patients \cite{mertens_differential_2015} and prevents herpes simplex outbreaks \cite{ferensztajn-rochowiak_long-term_2023, rybakowski_antiviral_2022}. \\

\begin{figure}
	\hspace{-10mm}
	\begin{minipage}{0.6\textwidth}
		\includegraphics[width=1\linewidth]{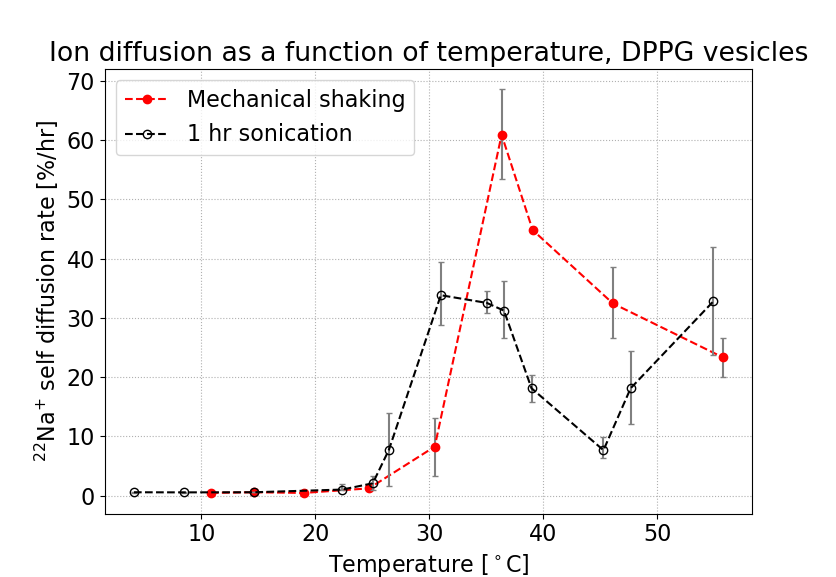}
		\caption{Sodium diffusion in DPPG bilayer vesicle suspensions as a function of temperature. The peak in diffusion is close to the melting temperature of the lipid vesicles in the range of 32-42$^\circ$C. Diffusion rates were not converted to permeability rates due to lack of sample homogeneity. Adapted from \citet{papahadjopoulos_phase_1973}.}
		\label{fig:Papa1973}
	\end{minipage}
	\hspace{10mm}
	\begin{minipage}{0.6\textwidth}
		\includegraphics[width=1\linewidth]{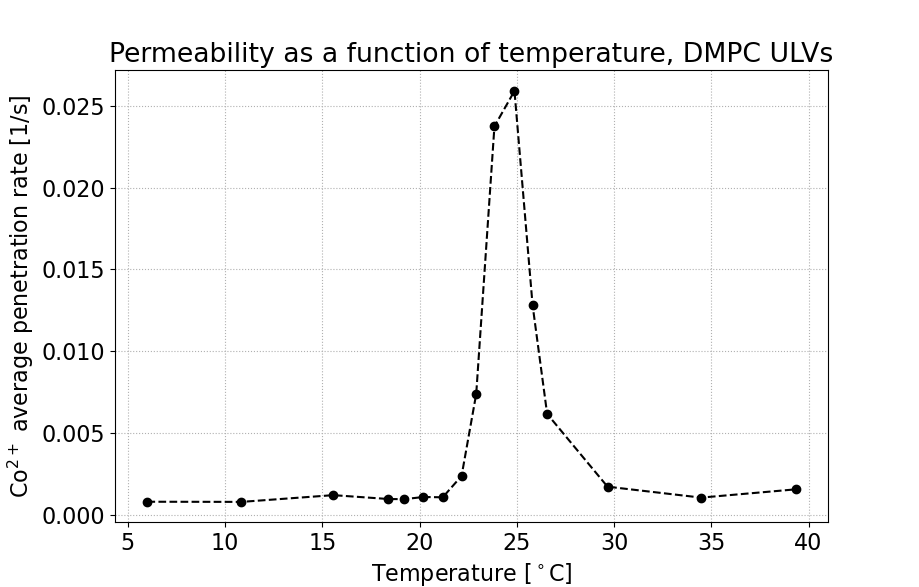}
		\centering
		\caption{Cobalt permeability in unilamellar DMPC vesicles as a function of temperature. Similarly to the figure \ref{fig:Papa1973} to the left, a peak is visble close to the melting temperature of the vesicles, with a midpoint at 24.7$^\circ$C. The average time constant is inversely related to time it takes for a cobalt ion to penetrate the membrane. Adapted from \citet{sabra_lindane_1996}.}
		\label{fig:Sabra1996}
	\end{minipage}
\end{figure}

\begin{figure}
	\centering
	\includegraphics[width=0.7\linewidth]{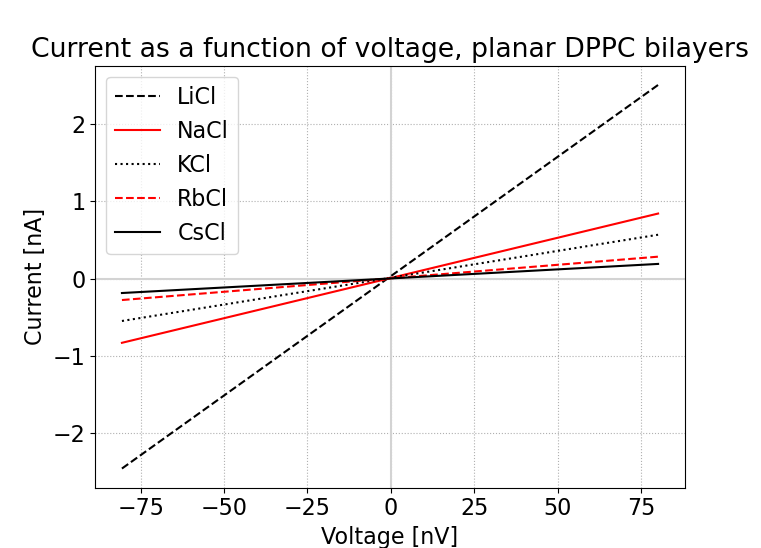}
	\caption{Current across DPPC lipid membranes as a function of voltage, at pure lipid melting temperature (43 $^\circ$C). The slope of the graph indicates membrane conductivity (from Ohms law). The presence of lithium ions notably enhances the conductivity of the lipid membrane compared to other alkali metal ions. Planar lipid bilayer membranes in aqueous suspension, DPPC (pH 6.9). Adapted from \citet{antonov_soft_2005}.}
	\label{fig:Antonov2005}
\end{figure}

	\noindent\textit{Permeability}\\
Phospholipid bilayers undergoing phase transition become permeable to ions due to pore formation in the lipid membrane \cite{papahadjopoulos_phase_1973, antonov_appearance_1980, antonov_soft_2005, heimburg_lipid_2010, blicher_temperature_2009} (see figure \ref{fig:Papa1973} and \ref{fig:Sabra1996}). Lithium conductance during membrane transition is especially high compared to other alkali ions \cite{antonov_soft_2005} (see figure \ref{fig:Antonov2005}). It is quite likely that a considerable part of transmembrane lithium movement in cells and mitochondria are governed by lipid phase changes. About 70\% of cellular lithium influx and a minor but variable part of the efflux happens through a dissipative leak that seems to respond to membrane potential. This leak is enhanced \textit{in vivo}, indicating hydration dependency \cite{ehrlich_lithium_1980}. Furthermore, lithium ions seems to 'free-ride' on a number of otherwise highly specific ion channels involving protons, sodium, and calcium  \cite{jakobsson_towards_2017, walkon_calcium_2022}. Lithium ions interact with calcium ions \cite{fossel_lithium-7_1985}, potentially affecting calcium-induced permeability, as seen through up-regulation of genes associated with regulating calcium permeability during lithium response  \cite{rohr_impact_2022, mccarthy_calcium_2016, ament_rare_2015, chen_transcripts_2014}, suggesting these transport mechanisms might also be partially lipid phase mediated. \\

Interestingly, approximately one-fourth of the bipolar population have an inheritable imbalance in cellular in- and efflux of lithium ions, causing intracellular lithium to accumulate \cite{ostrow_heritable_1978, pandey_abnormal_1977, dorus_lithium_1979}. The imbalance is caused by decreased sodium-lithium counter-transport flow \cite{ostrow_heritable_1978}. The counter-transport relies on the actively upheld sodium gradient and the large discrepancy in membrane affinity \cite{ehrlich_lithium_1980}. This counter-transport in increased in patients suffering from hypertension, suggesting a potential phase-mediated role of cholesterol \cite{canessa_increased_1980}.\\







	\noindent\textit{Membrane buffers}\\
Phospholipids offer numerous binding sites that can buffer the ion concentration in cells \cite{binder_effect_2002}, and the availability of these sites are majorly influenced by competitive ion binding, electrostatic screening from ion species in the surrounding environment, or changes in lipid conformation \cite{trauble_electrostatic_1974, lopez_cascales_effect_1997}. \\

\begin{figure}
	\centering
	\includegraphics[width=0.7\linewidth]{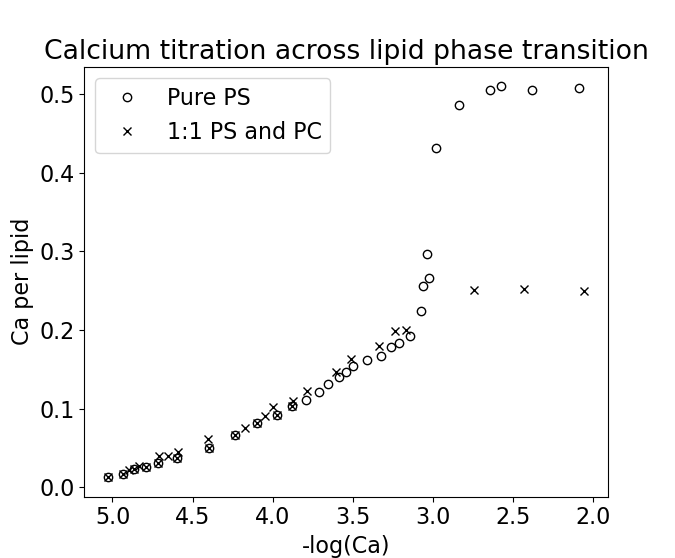}
	\caption{Titration of membrane-bound calcium in in unilamellar vesicles of pure DPPS dispersion versus 1:1 mix DPPS and DPPC. There is a discontinuous increase in membrane-bound calcium at phase transition (from fluid to gel phase with change in geometry), most notable in the case of pure PS. Data adapted from \citet{ekerdt_intermembrane_1982}.}
	\label{fig:Ekerdt1982}
\end{figure}

Buffer release caused by changes in lipid-phase is a surprisingly strong effect. Ion membrane affinity is higher in lipid gel phase than in fluid phase in both charged and zwitterion membranes (PS and PC) \cite{binder_effect_2002, shih_membrane_2018, ekerdt_intermembrane_1982}. \citet{ekerdt_intermembrane_1982} showed that changing the membrane phase from fluid to gel through calcium titration causes a rapid increase in membrane-bound calcium around the lipid bilayer melting point (see figure \ref{fig:Ekerdt1982}), while \citet{shih_membrane_2018} estimated the calcium-binding ability of DMPC nanodiscs to increase by 40\% from fluid to gel-phase. These percent wise changes are in the same order of magnitude as the changes seen in fig. \ref{fig:McCarthy2016}.\\

Lipid-phase mediated buffer signaling could take place in the following way: 
Increasing or decreasing the lipid melting point of the cell membrane by changing the lipid phase will cause the cell membrane to either take up or release ions in the intracellular environment. In the case of calcium signaling, the presence of calcium, and to less extent lithium, at the cell interface increases the melting point of an otherwise fluid cell membrane. This will increase the membrane calcium binding property, causing a transient uptake of intracellular ions, prompting intracellular buffering membranes (ER, mitochondria, and MAMs)  to release ions. A small but abrupt dip in intracellular calcium concentration prior to intracellular calcium release can actually be seen in figure \ref{fig:McCarthy2016}. Oppositely, melting point depression in the cell membrane would release ions into the cytosol, a mechanism that has previously been proposed for cell signaling with alkali-ions \cite{trauble_electrostatic_1974}.\\


But for this mechanism to work, the general membrane state has to be sufficiently close to the domain transition point. This might explain why the ongoing presence of lithium in the extracellular environment depletes calcium reactivity. Since lithium stabilizes the gel-state much better than other alkali ions, and this effect is apparent even in mixed ion solutions, its continuous presence could move the membrane state away from where it is the most responsive. The body would be able to counteract this by altering it's lipid expression, such that the membrane phase becomes more fluid, and thereby renormalize cellular calcium reactivity over time. \\

In this simple framework, the lipid membrane is assumed to generally be in the fluid state. In reality, the cell membrane has a complex expression of lipid species and proteins, and the melting temperature is not entirely uniform across the membrane. Some membrane domains have an increased melting point and may thus express a higher lipid buffer ability (potentially increased by the presence of buffering proteins), such as PS and PE domains in the inner leaflet of the red blood cell membrane  \cite{mclaughlin_adsorption_1981}. \\

A pathological offset in the membrane lipid state might be the reason why fibroblasts from bipolar patients do not yield as dramatic a calcium-release as control when subjected to acute lithium. In the section \textit{Applications to cellular function} we further discuss indications that there might be an offset in membrane phase in bipolar patients.\\

	\noindent\textit{Lithium's effect on proteins}\\
	Lithium associates to a plethora of proteins in the human body and have denaturing properties \cite{clarke_binding_2004, beauchamp_effect_2012}. Similarly to lipids, ions primarily interact with proteins and affect their conformation through hydration \cite{binder_effect_2002}, and ion association is likewise highly specific. Lyotropic series of lipids and proteins often resemble each other \cite{binder_effect_2002, pgregory_understanding_2022}, and lithium's deviating behavior can be seen in both. The ability for an ion species to precipitate protein in solution is described by the Hofmeister series, and here lithium, oppositely to other alkali ions and similarly to calcium, often increases solvation \cite{pgregory_understanding_2022}. \\

Proteins with exceptionally high affinity for divalent ions, specifically calcium, have been found in the mitochondria  and the sarcoplasmic reticulum \cite{hauser_interactions_1976}. The involved binding sites have been proposed by \citet{hauser_interactions_1976} to be carboxylate and phosphate groups, which in phospholipid membranes is the main target of protons and lithium ions. Lithium may screen these binding sites, affecting calcium buffer ability and possibly the proton transport involved in mitochondrial energy metabolism \cite{wang_energy_2023}. In fact, acute lithium has been shown to directly enhance oxidative phosphorylation respiratory chain activity in enzymes isolated from human mitochondria in a dose-dependent manner \cite{maurer_lithium-induced_2009}. Older experiments involving various enzymatic activity reviewed by \citet{schou_biology_1957} showed similar, activating effects. It is likely that a significant component of lithium response is concerned with readjusting the thermodynamic state of proteins that have been modified by the presence of lithium. This subject is highly relevant, but it is beyond the scope of this paper. \\

	\noindent\textit{Applications to cellular function}\\
Several fundamental biological processes have been connected to phase changes in the membrane lipid matrix. \\

One such example is the nerve signal. The many physiological changes associated with nerve pulse propagation \cite{heimburg_soliton_2005, andersen_towards_2009, mosgaard_timescale_2015, kamp_numerical_2021} are predicted by soliton theory to stem from the transient lipid phase changes observed in the nerve membrane during the action potential \cite{georgescauld_transient_1978,heimburg_important_2021, tasaki_volume_1990, fedosejevs_sharp_2022}. In this framework, the phase change is described as a soundwave propagating through the membrane. The changes in permeability, voltage, and physical dimensions follow from linear non-equilibrium thermodynamics. By including lipid phase change in the description of the nerve pulse, phenomena such as nerve pulses initiated by local cooling \cite{heimburg_action_2007, carr_effects_2003} and mechanical stimulation are accounted for \cite{heimburg_physics_2010, iwasa_mechanical_1980, kusano_mechanical_1990}. Soliton theory additionally offers a physical interpretation for the efficacy of drugs with lipid-phase altering properties, such as general anesthesia \cite{heimburg_action_2007, seeger_how_2007}. \\

In a previous work, the author investigated lithium therapy within this framework \cite{kamp_numerical_2021}. It was argued similarly that acute lithium exposure orders the lipid phase of the nerve membrane and brings it closer to transition. This increases the nerve's sensitivity to disturbances in the environment, since less work would be required to push the lipid phase into transition and start a propagating nerve signal. The nerve picks up on small changes in membrane phase induced by their surroundings, such as sub-threshold firing of neighboring neurons. This lithium-induced hyperexcitability of the nerve membrane is simultaneously associated with a decrease of calcium-buffer ability, possibly disrupting intercellular communication with mitochondria and other intercellular functions.\\

Inside the cell, several well-known phenomena are worth investigating for lipid-phase mediated effects. Cholesterol-rich lipid rafts and other types of micro-domains may utilize phase-mediated calcium buffering \cite{pani_lipid_2009, berridge_calcium_2006}. 
The mitochondrial permeability transition pore, a spontanous transition in permeability of the inner mitochondrial membrane associated with necroptosis  \cite{bauer_role_2020}, is initiated by an overload of calcium ions or reactive oxygen species, and prevented by the presence of lithium, indicating calcium-lithium interactions or lipid adaption \cite{deline_lithium_2023, shalbuyeva_lithium_2007}. The pore formation might be caused by lipid order-inducing compounds pushing the mitochondrial membrane phase far into the transition regime, causing an abrupt increase in permeability \textit{(See 'Permeability')}.\\

Lipid peroxidation and aging has been associated with a significant decrease in fluidity of the cellular membrane. Oxidative species alter the structure of lipid headgroups \cite{valvassori_lithium_2017} and decrease fluidity in pure lipid membranes and mitochondrial membranes \cite{borst_oxidation_2000, chen_alterations_1994}. The gel-phase is more vulnerable to oxidation than the fluid phase \cite{oteiza_mechanism_1994, verstraeten_effect_1997}, a possible explanation why long-term lithium therapy specifically protects against oxidative compounds \cite{lai_cytoprotection_2006}. Additionally, non-steroidal anti-inflammatory drugs (NSAIDs) disrupt lipid membrane order \cite{lichtenberger_insight_2012, sharma_effects_2020} and increase lithium serum values. \\
\indent Cellular aging is proposed to be caused by long-term peroxidation \cite{chen_alterations_1994}, and is associated with significant changes in membrane composition and dynamics \cite{skowronska-krawczyk_aging_2020}, such as an upconcentration of cholesterol in the outer layer of the cell membrane \cite{gibson_wood_cholesterol_2011} and an increased fluidity in calcium-buffering, intracellular platelet membranes,  the latter hypothesized to be due to a lack of membrane cholesterol \cite{van_rensburg_lipid_1994}. Since cholesterol attenuates the lipid phase transition, the increase in plasma cholesterol could help explain disrupted cellular signaling in old age for excitable tissues such as nerves or the heart, and a possible cause for the inverse relationship between cellular maturity and hyper-excitability in bipolar patients \cite{mertens_differential_2015, dmello_lithium_1994}.\\
\indent Alzheimer's disease, a pathological version of aging, has similarly been associated with increased plasma membrane cholesterol \cite{loera-valencia_alterations_2019, marquer_increasing_2014} and increased fluidity of intracellular platelet membranes, especially the endoplasmic reticulum \cite{van_rensburg_lipid_1994}. Interestingly, lithium treatment has shown to be effective in preventing cognitive decline in Alzheimer's disease and mild cognitive impairment \cite{matsunaga_lithium_2015}. It is possible that the redistribution of cholesterol during aging is at the expense of intracellular cholesterol, and that this can be partially counteracted by lithium-induced phospholipid- and cholesterol production.\\

The mechanisms in lithium response central to bipolar disorder, namely changes in circadian rhythm, lipid expression and peroxidation levels, mitochondrial function, cellular calcium reactivity, and nerve activity, might hold crucial insight into pain reduction in lithium-responsive headaches. Abnormal lipid peroxidation and mitochondrial dysfunction have previously been documented in certain headaches, especially migraines \cite{alloush_mitochondrial_2019, sangiorgi_abnormal_1994, gross_mitochondrial_2021, wang_energy_2023}.\\


\section{An old and new perspective on lithium therapy}

	\noindent\textit{Bipolar homeostasis}\\
	From the framework of the previous section, we can infer a plausible cause for the hyperexcitable and erratic nerve function in bipolar patients. Overall, the state of their nerve membranes is simply too ordered, and thereby too far into lipid transition. There might be many initial causes: Wrong expression of membrane constituents, overproduction of 'activating' agents or, conversely, lack of relaxing agents, lowering the membrane melting point. It may even be a question of a single event of strenous cellular stress causing an unhealthy feedback cycle of increased lipid peroxidation and pH. \\

Introducing lithium would seem counterintuitive at first, since lithium ions would exacerbate this condition, not counteract it. But, as the body takes action, lipid phase becomes disordered and moves away from transition. The membrane phase is less sensitive and the membrane buffers at optimal level. This is a universal cellular adaptation for most living tissue, but, for some reason, only partially present in the bipolar population. This indicates that bipolar disorder is to some extent a lipid-metabolism disorder. \\

Diurnal periodicity could be of a similar nature as the nerve signal. A slow 24-hour variation of cellular membrane phase could lower cellular response and activity level at night-time and raise it during the day-time, as it is reported in literature \cite{allen_membrane_2017, emery_circadian_2012}. How the lipid matrix phase shift would affect the amplitude and period of such a system is currently not known, and we can therefore not say for sure that the period elongation caused by lithium response is due to melting point depression. Some indication is found in the circadian regulation of body temperature and thyroid stimulating hormone (TSH), where studies show disruptions in nocturnal and mean daily body temperature during affective episodes. Studies have found elevated nocturnal and mean-day temperatures in depressed patients \cite{souetre_twenty-four-hour_1988, mason_elevated_2024, avery_nocturnal_1982} and in patients receiving lithium therapy \cite{ mellerup_lithium_1978}, while manic patients exhibit large fluctuations and even episodes of decreased temperatures \cite{pflug_manic-depressive_1981, nikitopoulou_change_1976}. The decrease in circadian period observed in the lithium-responsive bipolar population at a low dose of lithium \cite{mccarthy_chronotype_2019} might in fact be lithium ions' direct impact on the biological clock, melting point elevation, in this special case unconcealed due to an impaired metabolic response. In this paper we have drawn a loose parallel between the period elongation of the diurnal rhythm of \textit{in vitro} neurons and fibroblasts to slowed heart rates in patients undergoing lithium treatment and anesthesia, but this is still speculative.\\


	\noindent\textit{Lithium withdrawal effects and drug interactions}\\
	What would then happen if we were to remove lithium from a lithium-responsive subject having undergone long-term lithium therapy? The ions may leave the body much faster than membrane lipid composition has time to adapt to. The balance between the dehydrating, solidifying presence of the ions and the adapted membrane would therefore be unbalanced. Initially, the membrane would thus be \textit{more} fluid than optimal, and the melting point of the nerve membrane would be \textit{depressed}. The subject would be expected to feel sedated for a few days as the body catches up. As the membrane re-normalizes, a bipolar patient would regain their hyperexcitable, manic state. \\

Melting point depression in the cell membrane likely takes place in many pharmacological treatments. It is especially well-documented in general anesthesia, where it has been proposed as the primary mechanism of action \cite{heimburg_action_2007, seeger_how_2007}. Anesthesia mixes well into the lipid membrane, disrupting order and changing lipid conformation. Anesthetic potency is tied very strongly to this ability \cite{heimburg_action_2007}. The acute effect of anesthesia and lithium are thus directly opposite, and presumably the same goes for the long-term adaptation. Long-term lithium administration has been shown to counteract the period-shortening effect of long-term ethanol administration in activity cycles of mice \cite{nascimento_alcohol_2015}. Since anesthesia is given short-term and lithium long-term, their effect would generally appear to be additive in literature, which is in fact the case \cite{leone_anesthetic_1984}. Additionally, anesthesia slows the heart similarly to long-term lithium treatment \cite{ilala_prevalence_2023, zanaboni_chapter_2007}. \\

Lithium has also been shown to interact with the cardiac glycoside ouabain. Lithium influx, but not efflux, into red blood cells is reduced by ouabain \cite{duhm_studies_1977}, indicating a potential impact on the leakiness of the lipid membrane. The effect of acute ouabain is also similar to lithium: single, low doses of ouabain induce hyperactivity in rats' brain tissue and in their behavior, while large doses are detrimental to function \cite{li_lithium_1997, schmerder_animal_2023}. In humans ouabain toxicity leads to increased heartbeat, twitching of muscles, convulsions, and death. Chronic lithium treatment increases the tolerance to ouabain, as seen through behavioral studies, tissue activity, and cell cultures \cite{schmerder_animal_2023, li_lithium_1997, el-mallakh_ouabain_2000}. Interestingly, the amount of ouabain binding to the membrane increases with chronic lithium treatment and could be an indicator of a partially lipid-mediated ouabain binding, which has previously been proposed \cite{herbette_structural_1986}. \\  


	
	\noindent\textit{Bipolar disorder and alcohol abuse}\\
	A single but unfortunate example of long-term use of general anesthetics is alcohol abuse. Ethanol is an anaesthetic, albeit a less effective one, and it is likely not a coincidence that there is a strong history of alcohol abuse with bipolar disorder \cite{grunze_comorbid_2021}. Following the reasoning presented in this paper, acute alcohol consumption might transiently move the bipolar nervous system away from its proximity to lipid phase transition and thereby ease neuronal activity. But, unfortunately, the long-term adaptation to chronic alcohol consumption would counteract this by increasing the melting point. Without the constant presence of alcohol, the patient is worse off than before, trading manic symptoms for the tremor, agitation, seizures, and delirium of alcohol withdrawal \cite{saitz_introduction_1998}. In this framework, manic delirium, delirium tremens, and lithium intoxication are all expressions of the nerve membrane phase being out of balance and therefore similar in nature. \\

\section{Closing remarks}

	Lithium therapy has now been divided into three components: the preexisting membrane state, the acute effect of lithium ions, and the long-term adaptation by the body. \\

This should be especially useful in drug development. Pathological imbalances in cellular membrane state can, in early stages of development, be represented by an \textit{in vitro} lipid vesicle model with careful consideration of environmental conditions. One would then have to make a strategic decision whether to develop drugs for immediate relief by directly counteracting the pathological offset, or more ambitiously attempting to induce a reversal of the offset through a metabolic response similar to what is seen lithium therapy. The first approach might lead issues with long-term tolerance as the body adapts to the presence of the drug, while efficacy of the latter entirely depends on the patient's ability to produce the correct response, preferably not at the detriment of other biological functions. One can also impact membrane phase through non-pharmacological means, such as specialized diets \cite{sethi_ketogenic_2024, tandon_effect_1998} and heat training \cite{mishra_circadian_2021, mischoulon_heat_2017, tsagkaris_when_2022}.\\

It is a big caveat of therapies based on membrane phase regulation that treatment is chronic and needs careful monitoring. If the therapy fails to maintain the optimal membrane state perfectly, it might be detrimental to the patient. Outer influences, such as dehydration or altered salt intake, can lead to acute lithium toxicity in an otherwise stabilized patient on lithium therapy. Ideally, one would, in the future, be able to target the underlying illness by pinpointing the genetic abnormalities or the broader mechanism responsible for the atypical membrane state and offer specialized therapy.\\

One could aim at improving pre-existing therapies by investigating their lipid phase-altering properties, such as antiviral medication, antibiotics \cite{seeger_how_2007, ermishkin_single_1976, kofman_inhibition_1990}, neurotransmittors \cite{seeger_how_2007}, antidepressants \cite{fisar_effect_2005, birkinshaw_antidepressants_2023, patetsos_treating_2016}, and the neuroactive compound tetrodotoxin \cite{agui-gonzalez_secondary_2021}. Lithium therapy itself might be even improved. Studies of animal behavior and mitochondrial permeativity indicate that there might be a therapeutic difference between the two stable lithium isotopes, Li-6 and Li-7 \cite{sechzer_aberrant_1986, li_nuclear_2018, deline_lithium_2023, livingstone_search_2023}. Li-6, of only 7.5\% abundance might be more potent than the more prevalent Li-7 (92.5\% abundance) \cite{deline_lithium_2023}. Isotope effects are well-known within quantum chemistry and are most prominent at low mass, where small changes in hydration shells of ions can greatly affect solvation free energy. The diffusion constant in water have been shown to differ between Li-6 and Li-7 \cite{renshaw_diffusional_1987}. \\

This paper demonstrates that the long-disputed fundamental mechanism of lithium therapy may be quite elementary, and serves as another example of phase regulation in the body. The body's response is likewise straightforward in its purpose, though the underlying metabolic pathways are highly complex.\\

	
	\subsection{Acknowledgments}
	I would like to thank my supervisor, Karel Proesmans, for supporting me on this project, Thomas Heimburg for ongoing motivation and good discussions, Lily Thijssen for guidance and support, and Dan Farbowitz for editing and proof-reading.\\

This work was supported by the Novo Nordisk Foundation [NNF18SA0035142, NNF21OC0071284].

\bibliographystyle{unsrtnat}
\bibliography{references_R2}
\end{document}